 \documentclass[final,5p,times,twocolumn,number]{elsarticle}

\usepackage[parfill]{parskip}
\usepackage[skip=5pt]{caption} 
\usepackage{amssymb}
\usepackage{lipsum}
\usepackage{graphicx}
\usepackage{booktabs,siunitx}
\graphicspath{{images/}}
\usepackage[hyphens]{url}
\usepackage{hyperref}
\usepackage{xurl}

\journal{arXiv}

\begin{document}

\begin{frontmatter}



\title{Diversity Over Scale: Whole-Slide Image Variety Enables H\&E Foundation Model Training with Fewer Patches}


\author[paicon]{Christoph Bosch}
\author[cloud]{John K.L. Wong}
\author[atb]{Martin Paulikat}
\author[cloud]{Myroslav Zapukhlyak}
\author[cloud]{Bharti Arora}
\author[paicon]{Manasi Aichmüller-Ratnaparkhe}
\author[paicon]{Jens Baumann}
\author[ixora]{Shivani Karn}
\author[ixora]{Rutuja Kamble}

\author[megavision]{Swapnil Karnik}
\author[megavision]{Bhushan Khedkar}

\author[chhut]{Serey Vathana Chhut}

\author[paicon]{Witali Aswolinskiy}

\author[paicon,cloud]{Christian Aichmüller\corref{cor1}}

\affiliation[paicon]{organization={Paicon GmbH},%Department and Organization
            city={Heidelberg},
            country={Germany}}
\affiliation[cloud]{organization={Paicon Cloud GmbH},%Department and Organization
            city={Heidelberg},
            country={Germany}}            
\affiliation[atb]{organization={Department of Applied Tumor Biology, Universitätsklinikum Heidelberg},%Department and Organization
            city={Heidelberg},
            country={Germany}}
\affiliation[ixora]{organization={IXORA Digital Health Private Limited},
            city={Pune},
            country={India}}
\affiliation[megavision]{
    organization={Mega Vision Labs},
    city={Pune},
    country={India}
}
\affiliation[chhut]{%
    organization={University of Health Science},
    city={Phnom Penh},
    country={Cambodia}
}

\cortext[cor1]{Corresponding author: c.aichmueller@paicon.com}

\begin{abstract}
Rapid progress in computational pathology is increasingly driven by vision foundation models pretrained on vast histopathology datasets. While recent efforts have prioritized  training on an ever-larger amount of patches, we take an alternative approach focused on data diversity. Our foundation model, Athena, was initialized from a pretrained model and trained on just 115 million tissue patches, several times fewer than recent histopathology foundation models. Rather than relying on patch volume or complex sampling heuristics, we maximize data diversity by randomly selecting only a moderate number of patches per whole-slide image from our diverse internal repository, which spans multiple countries, institutions, and scanner types.
Evaluated on a single patch-level benchmark and four slide-level downstream tasks (two molecular and two morphological), Athena approaches the state-of-the-art and even surpasses several models trained on substantially larger datasets. This indicates that diversity across whole-slide images, rather than patch quantity alone, drives learning in histopathology foundation models.

\end{abstract}



%% \begin{keyword}
%% keywords here, in the form: keyword \sep keyword, up to a maximum of 6 keywords
%% keyword 1 \sep keyword 2 \sep keyword 3 \sep keyword 4



%% \end{keyword}

\end{frontmatter}

%\tableofcontents



\section{Introduction}
%%\label{}
Digital histopathology has transformed cancer diagnostics, providing a foundation for precision oncology and adaptive, patient-tailored treatment strategies. Among its modalities, hematoxylin and eosin (H\&E) stained whole-slide images (WSIs) are pivotal for diagnosis, prognosis, and therapy planning in cancer, where microscopic tissue evaluation reveals pathological features.  Recently, interest has grown in developing vision foundation models (FMs) specific to computational pathology. This is because the features learned by such models can be used for many different downstream applications like cancer subtype prediction, molecular classification \cite{giga}, survival prediction \cite{survival} or treatment outcome \cite{treat}.

The primary drivers of the recent advancements are the availability of numerous WSI datasets (public and proprietary), rapid progress in architectures for large-scale vision models \cite{vit}, and improved self-supervised training frameworks for large datasets like DINOv2 \cite{dino}.
DINOv2 employs a self-distillation approach where a teacher model guides a student model to match embeddings across augmented crops, enabling ViT encoders to capture both global context and local structure for general-purpose vision FMs.
Recent state-of-the-art histopathology foundation models are predominantly based on large-scale DINOv2 pretraining, leveraging datasets of up to 3.4 billion patches derived from over one million slides \cite{atlas}.

In this work, we examine whether comparable performance can be achieved using far fewer training patches. We demonstrate that this is possible through several efficiency-oriented design choices: 1.) initializing from a pretrained model, 2.) adopting the DINOv2 training pipeline with only minimal modifications, and 3.) employing a random patch-selection strategy across a highly diverse set of WSIs.

\section{Methods}
%%\label{}

\subsection{Training data}

Our WSI repository comprises a diverse collection of slides sourced from centers worldwide and digitized using eight different scanner models from major vendors. These slides exhibit substantial variability in staining, scanning protocols, organ types, and tissue preparation, making them well suited for training a robust and generalizable FM.
For training Athena, we used 274,150 H\&E whole-slide images from our repository, along with 8,400 slides from  GTEx \cite{gtex}.

\begin{figure}[h!]
  \centering
  \includegraphics[width=.4\textwidth, ]{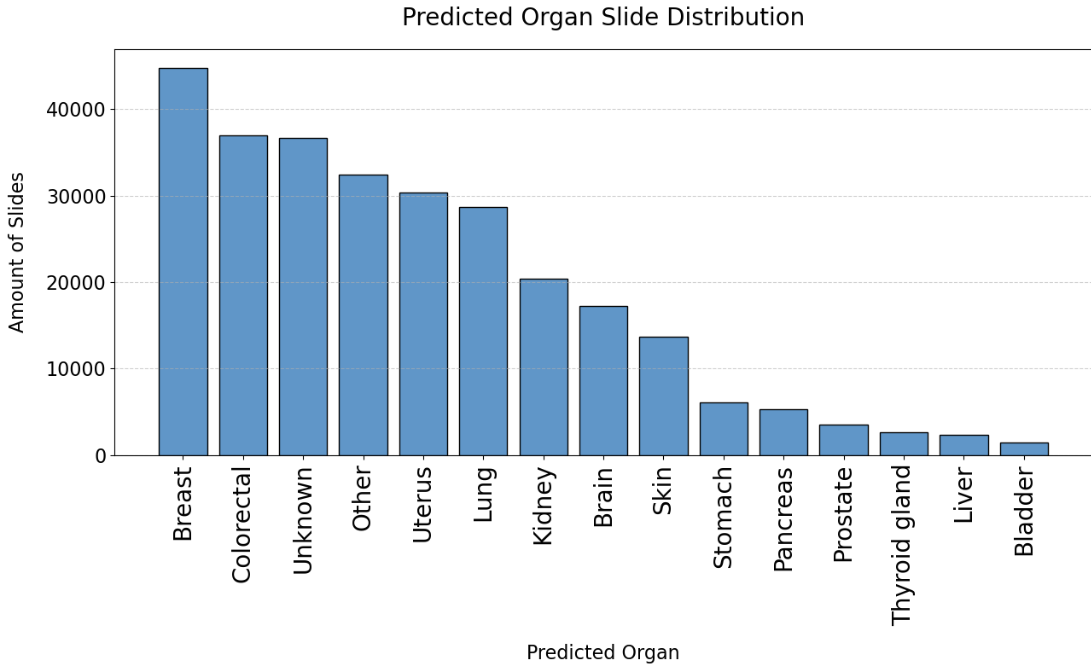}
  \caption{Organ distribution across slides}
  \label{fig:organ_distri}
\end{figure}

An overview of the dataset composition is shown in Figure~\ref{fig:organ_distri}, illustrating the distribution of organs across all training slides. Organ labels were generated using the organ classifier proposed in \cite{organ_classifier}. The "Other" category includes slides from organ types not covered by the classifier, while "Unknown" denotes slides known to be of human origin and H\&E stained, but for which no classification results are currently available.

To extract the training patches, we first applied a tissue filter mask to remove background regions. The mask was created using a combination of a contrast-based and a LAB color space–based threshold to isolate tissue regions. From the remaining tissue areas, we randomly sampled approximately 410 patches per slide across four resolutions (0.25, 0.5, 1.0, and 2.0 µm per pixel), allocating twice as many patches at spacing 0.5 and 1.0 as at 0.25 and 2.0. We excluded blurry patches using a Laplacian filter and removed low-information patches using the HSV-based method described in \cite{midnight}. To mitigate overrepresentation from any single country, patch sampling was adjusted to partially balance dataset size disparities across the 24 contributing countries. Specifically, patches from slides belonging to underrepresented countries were upsampled (by up to a factor of four) to improve the geographic balance of the training set. This process resulted in a total of approximately 115 million patches used for model training.

\begin{figure}
  \centering
  \includegraphics[width=.4\textwidth, angle=270]{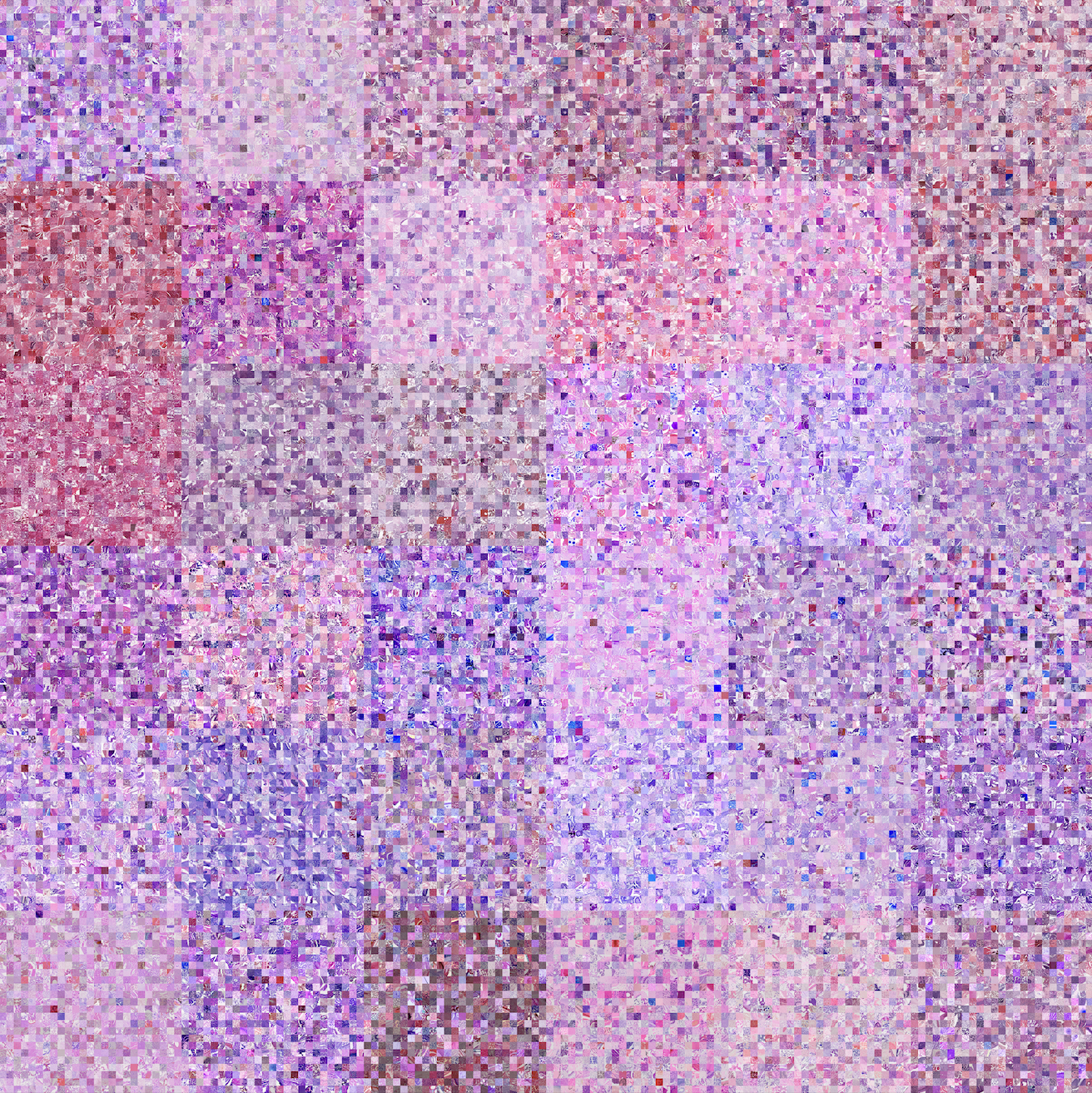}
  \caption{Visualization of staining diversity: The 36 (6×6) quadrants are generated from 25×25 randomly sampled patches per cohort, showcasing the heterogeneity introduced by different scanners, staining methods, and tissue preparation procedures.}
  \label{fig:visualize_buckets}
\end{figure}

To illustrate the diversity in our dataset, Figure~\ref{fig:visualize_buckets} shows the variation in H\&E staining across different centers. Each of the 36 quadrants (consisting of 25x25 random sampled patches) represents one cohort, highlighting how differences in scanners, staining protocols, and tissue preparation result in distinct color profiles. This underscores the importance of training models on diverse slides to ensure robustness to such variability.

\begin{figure}[h!]
  \centering
  \includegraphics[width=.5\textwidth]{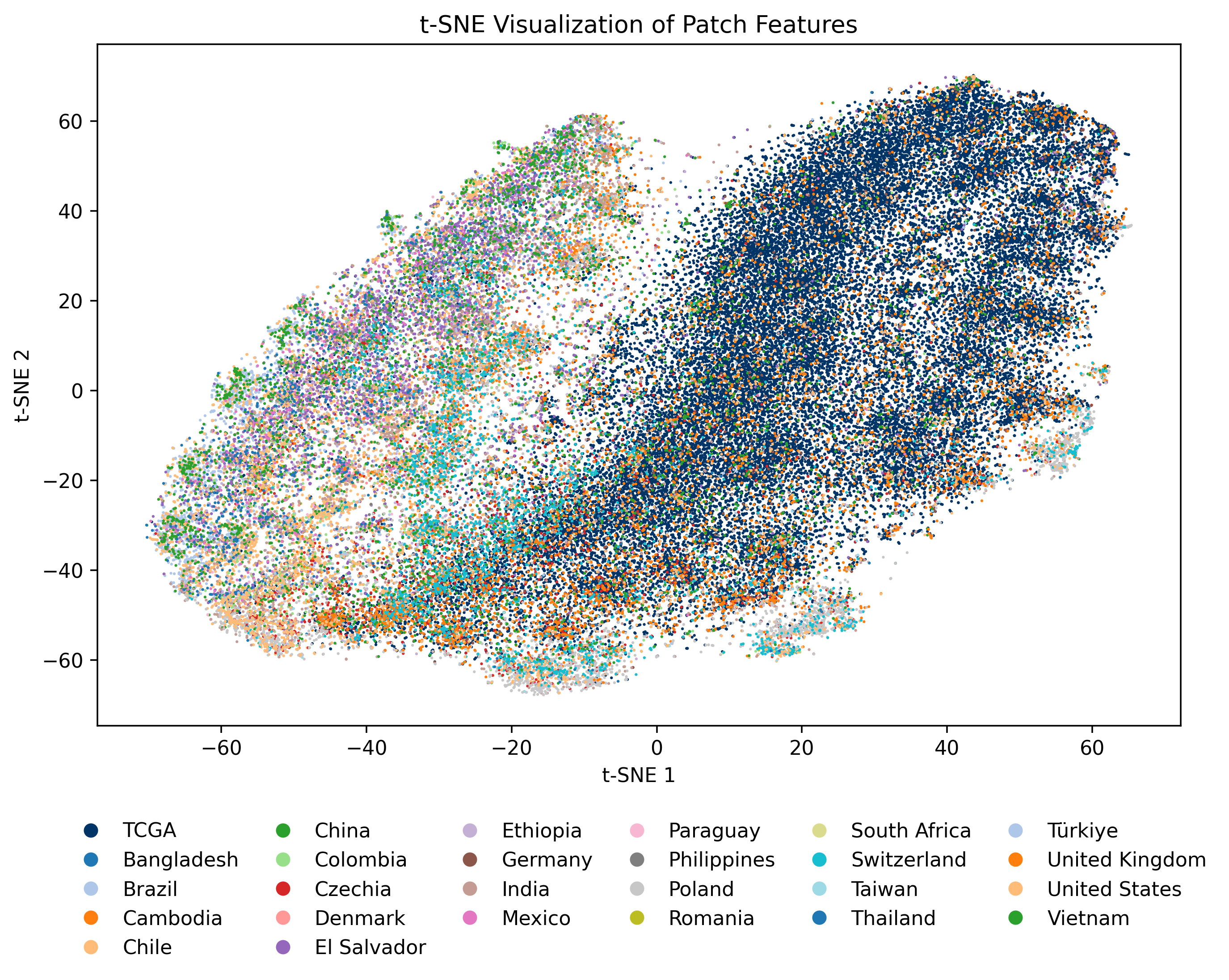}
  \caption{t-SNE visualization of patch features from TCGA \cite{tcga} and Athena training slides}
  \label{fig:tsne}
\end{figure}

To further visualize the patch diversity of our dataset, we generated a t-SNE \cite{tsne} plot using 40k patches from TCGA \cite{tcga} and 40k patches from different countries of our training dataset (Figure~\ref{fig:tsne}). Four patches per slide were extracted at a spacing of 0.5 µm per pixel and each patch was represented by 2×2 average-pooled, normalized raw RGB features. The resulting t-SNE visualization reveals overlapping distributions between TCGA and our dataset in some parts. However, several country-specific regions are uniquely represented in our data. This indicates that our dataset contains a broader range of feature variability, likely reflecting country-specific variation arising from differences in staining protocols, scanner characteristics, and other technical factors.

\subsection{Self-supervised training with DINOv2}
Following recent FMs in histopathology, we adopt DINOv2 \cite{dino} as our training framework. A 1.1-billion-parameter ViT-G \citep{vit} was trained in a self-supervised setup on 115 M patches. To reduce training time, we initialized the model from a DINOv2-pretrained ViT-G trained on natural images (LVD-142M \cite{dino}). Training was conducted on four nodes with eight H200 GPUs each, resulting in a global batch size of 2880. The overall training time was 18 hours (576 GPU-hours).
To maximize batch diversity, each batch was constructed using patches randomly sampled from across all slides rather than from a single slide.

Our setup closely follows the original DINOv2 configuration, with minor modifications to the learning rate schedule and augmentations. We applied a 20 k-iteration linear warm-up of the learning rate to \(8\cdot 10^{-4}\), which remained constant thereafter. A vertical flip (probability = 0.5) was added to the augmentations to reflect the orientation invariance of histopathology images. The model was trained for 40 k iterations (115 million patch views), during which the total training loss decreased from 14 to 6.8.

\subsection{Multiple Instance Learning}

For the evaluation on slide-level tasks, the procedure is as follows.
First, background regions are removed from each slide using the same thresholding method applied during the training patch extraction. The remaining tissue is divided into non-overlapping 224×224 patches and encoded using the FM by extracting the class token. These encodings are then used to train attention-based multiple instance learning (MIL) networks, following the general formulation of \cite{ilse} and \cite{clam}, but with a simplified non-gated attention mechanism, resulting in a lightweight and interpretable aggregation of patch features for slide-level prediction.

\subsection{Evaluation}

Benchmarking of foundation models in histopathology remains an open challenge, as no standardized evaluation protocols have yet been established despite ongoing community efforts.
We benchmark Athena against other FMs using a single patch-level task, HEST \cite{hest}, and four slide-level downstream tasks: two morphological (Breast IDC/ILC \cite{tcga}, CAMELYON16: Lymph node Tumor/Normal \cite{cam16}) and two molecular (MSI/MSS \cite{kather}, HER2 positive/negative \cite{herohe}).

For all slide-level tasks except CAMELYON16, cross-validation was employed to create an ensemble of 40 MIL-models, ensuring stable results on the test datasets. Five folds were used for HER2 positive/negative and IDC/ILC, and four folds for MSI/MSS. For CAMELYON16, we trained 40 models with different random seeds and report the average performance across runs. The models were trained on spacing 1 for MSI, HER2 and IDC/ILC and spacing 0.5 for CAMELYON16.

Table~\ref{tab:model-comparison} provides an overview of the evaluated models and their corresponding training data volumes.
The training runs for Athena and Midnight-12 were initialized from a pretrained model on natural images, whereas all other models were trained from scratch.
Although UNI \cite{uni} was trained on 100 million unique patches, the model was exposed to a total of 384 million patch views during training. Athena was trained on the smallest total number of patches among all models. 

\begin{table}[h!]
\centering
\small                                % one step bigger
\setlength{\tabcolsep}{6pt}           % default-ish column spacing
\renewcommand{\arraystretch}{1.15}    % a bit taller rows
\caption{Comparison of histopathology FM models}
\label{tab:model-comparison}
\begin{tabular}{l r r l}
\toprule
\textbf{Model} & \textbf{WSIs} & \textbf{Patches} & \textbf{Architecture} \\
\midrule
Athena        & 282k  & 115M  & ViT-G/14 \\
H-optimus-0 \cite{hopt0}   & 500k+ & 100M+ & ViT-G/14 \\
Midnight-12 \cite{midnight}      & 12k   & 384M  & ViT-G/14 \\
Virchow \cite{virchow}       & 1.5M  & 2.0B  & ViT-H/14 \\
Hibou-L \cite{hibou}        & 1.1M  & 1.2B  & ViT-L/14 \\
UNI \cite{uni}          &  100k  & 100M & ViT-L/14 \\
\bottomrule
\end{tabular}
\end{table}

\subsubsection{HEST Benchmark}

The HEST benchmark \cite{hest} comprises nine tile-level regression tasks designed to evaluate a model’s ability to predict gene expression from histology images. For each task, embeddings of 224×224 tiles at 0.5 µm per pixel are regressed against normalized transcript counts of the 50 most variable genes measured at the corresponding spatial locations. Performance is quantified by the Pearson correlation between predicted and actual gene expression, aggregated across patients.
Following \cite{midnight}, we concatenate the class token with the mean of all patch tokens for HEST, which yields better average performance than using the class token alone.

\subsubsection{MSI vs. MSS Classification}
Microsatellite instability (MSI) status is an important biomarker in colorectal cancer, influencing both prognosis and therapeutic decisions \cite{kather}. We trained our models on 421 TCGA \cite{tcga} and 221 CPTAC \cite{cptac} slides, and evaluated their performance on independent test sets comprising 47 PAIP \cite{paip}, 250 NIB \cite{nib}, and 850 SURGEN \cite{surgen} slides.

\subsubsection{HER2 Positive vs. Negative Classification}
Accurate assessment of HER2 status is critical for breast cancer subtyping and directly impacts both prognosis and treatment planning. Our models were trained on 971 TCGA \cite{tcga} and 323 CPTAC \cite{cptac} slides, and evaluated on diverse test sets, including 191 YALE-HER2 \cite{yaleher2}, 508 HEROHE \cite{herohe}, 1042 BCNB \cite{bcnb}, 126 IMPRESS \cite{impress}, 413 ACROBAT \cite{acrobat}, and 172 (proprietary) IXORA slides.

\subsubsection{CAMELYON16}
For the CAMELYON16 dataset \cite{cam16}, we followed the provided data split, using 270 slides for training and 129 for testing, without applying cross-validation. The task involves detecting breast cancer metastases in lymph node slides.

\subsubsection{IDC vs. ILC Classification}
We utilized breast cancer cases from the TCGA dataset \cite{tcga} to distinguish between invasive ductal carcinoma (IDC) and invasive lobular carcinoma (ILC). In total, 868 slides were used, with the test set comprising 50\% of the cohort (360 IDC and 74 ILC cases).

\section{Results}
%%\label{}
\subsection{HEST Benchmark}

\begin{table}[h!]
\centering
\small                                % one step bigger
\setlength{\tabcolsep}{6pt}           % default-ish column spacing
\renewcommand{\arraystretch}{1.15}    % a bit taller rows
\caption{Average Pearson correlation in HEST benchmark}
\label{tab:hest}
\begin{tabular}{l r}
\toprule
\textbf{Model} & \textbf{Score} \\
\midrule
H-optimus-0  & 0.425$\pm$.037 \\
Athena     &  0.417$\pm$.042 \\
Midnight-12   & 0.412$\pm$.036 \\
Virchow    & 0.406$\pm$.041 \\
Hibou-L      & 0.397$\pm$.052 \\
UNI        & 0.391$\pm$.049 \\
\bottomrule
\end{tabular}
\end{table}

Table~\ref{tab:hest} shows the results of the HEST benchmark for the different FMs, the results for Midnight-12, UNI, and Hibou-L were obtained from \cite{midnight}.
H-optimus-0 achieved the highest score, indicating the best overall performance among the evaluated models, followed closely by Athena. UNI showed the lowest performance. As the average standard deviation across tasks exceeds the differences between model scores, the results should be interpreted with caution, since the observed ranking may not be meaningful.

\begin{table}[h!]
\centering
\small
\setlength{\tabcolsep}{8pt}
\renewcommand{\arraystretch}{1.15}
\caption{Detailed HEST performance for Athena and H-optimus-0}
\label{tab:hest_detaill}
\begin{tabular}{l rr}
\toprule
\textbf{Dataset} & \textbf{Athena} & \textbf{H-optimus-0} \\
\midrule
READ       & 0.214$\pm$.057 & \textbf{0.240$\pm$.031} \\
IDC        & 0.593$\pm$.088 & \textbf{0.611$\pm$.080} \\
PRAD       & \textbf{0.375$\pm$.025} & 0.362$\pm$.021 \\
PAAD       & 0.490$\pm$.066 & \textbf{0.511$\pm$.047} \\
SKCM       & 0.636$\pm$.033 & \textbf{0.661$\pm$.055} \\
COAD       & \textbf{0.328$\pm$.016} & 0.309$\pm$.000\\
CCRCC      & 0.262$\pm$.049 & \textbf{0.292$\pm$.036} \\
LUNG       & \textbf{0.586$\pm$.002} & 0.575$\pm$.025 \\
LYMPH IDC & \textbf{0.268$\pm$.042} & 0.266$\pm$.039 \\
\bottomrule
\end{tabular}
\end{table}

Table~\ref{tab:hest_detaill} presents the detailed results for the two best-performing models, Athena and H-optimus-0. Overall, their performance is comparable across most tasks, with the largest differences observed in READ and SKCM (where H-optimus-0 performs better) and COAD (where Athena performs better).

\subsection{MSI vs. MSS Classification}

\begin{figure}[h!]
  \centering
  \includegraphics[width=.5\textwidth]{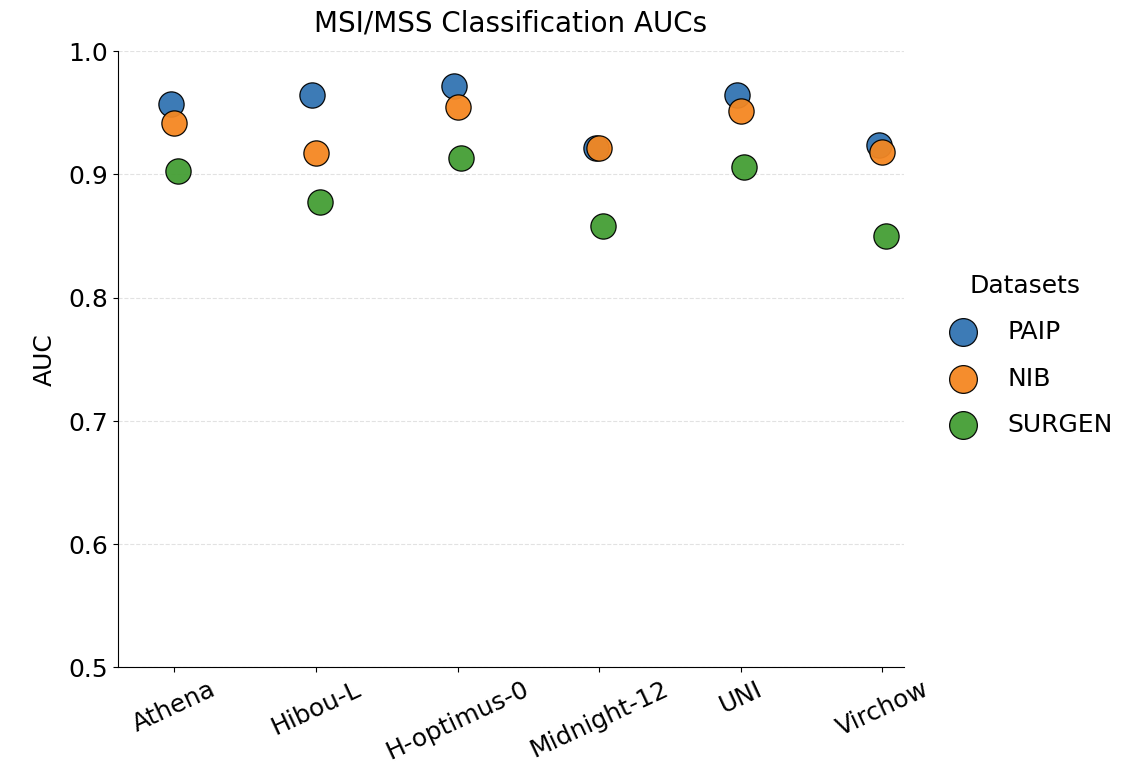} % no leading slash, no extension
  \caption{MSI/MSS classification AUCs for the different FMs}
  \label{fig:msi}
\end{figure}

Figure~\ref{fig:msi} and table~\ref{tab:msi} show the performance of the MSI/MSS classification. H-Optimus-0 consistently achieves the highest AUC across all datasets, with UNI and Athena showing comparable but slightly lower performance. Athena outperforms UNI and Virchow on all datasets and also surpasses Hibou-L on NIB and SURGEN.

\begin{table}[h!]
\centering
\small
\setlength{\tabcolsep}{6pt}
\renewcommand{\arraystretch}{1.15}
\caption{MSI/MSS classification AUC by model and dataset}
\label{tab:msi}
\begin{tabular}{l r r r}
\toprule
\textbf{Model} & \footnotesize\textbf{PAIP} & \footnotesize\textbf{NIB} & \footnotesize\textbf{SURGEN} \\
\midrule
Athena & 0.957 & 0.942 & 0.903 \\
Hibou-L & 0.964 & 0.917 & 0.878 \\
H-optimus-0 & \textbf{0.971} & \textbf{0.955} & \textbf{0.913} \\
Midnight-12 & 0.921 & 0.921 & 0.858 \\
UNI & 0.964 & 0.951 & 0.906 \\
Virchow & 0.924 & 0.918 & 0.850 \\
\bottomrule
\end{tabular}
\end{table}

\subsection{HER2 Positive vs. Negative Classification}

\begin{figure}[h!]
  \centering
  \includegraphics[width=.5\textwidth]{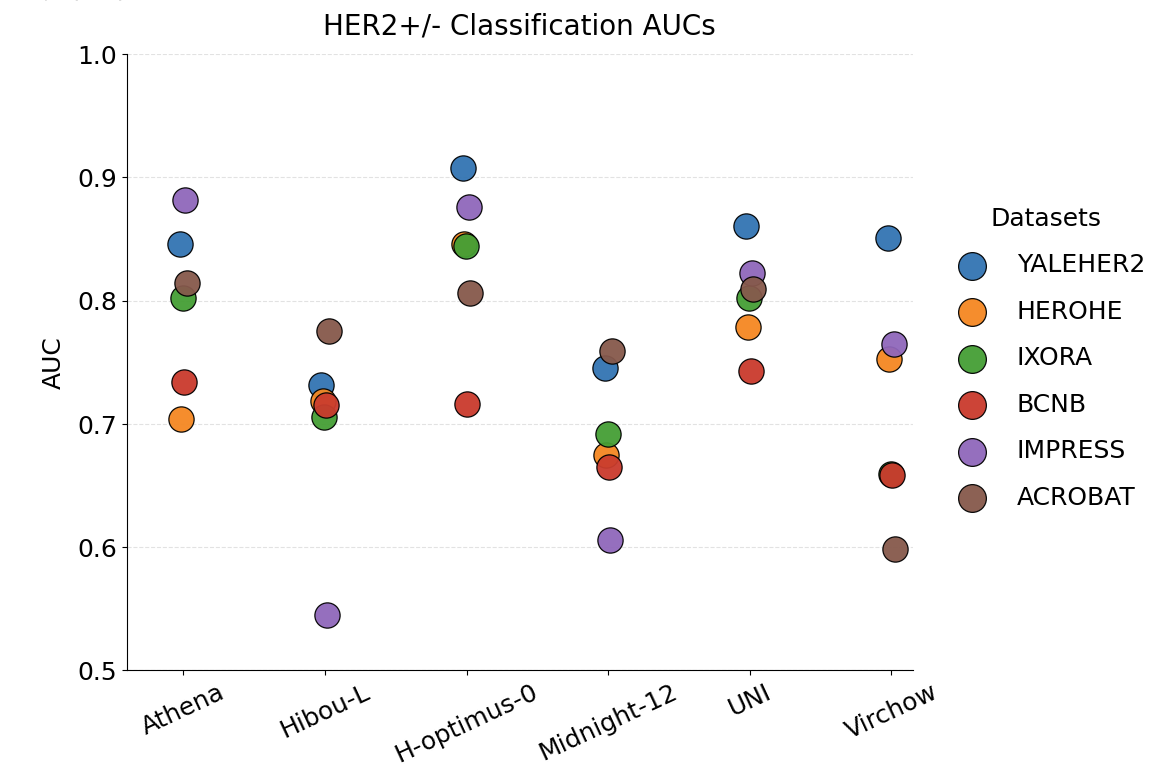} % no leading slash, no extension
  \caption{HER2 classification AUCs for the different FMs}
  \label{fig:her2}
\end{figure}

For HER2 classification, H-optimus-0 achieves the highest average performance but does not lead on every dataset, as shown in Figure~\ref{fig:her2} and Table~\ref{tab:her2}. The BCNB dataset appears to be the most challenging, likely because it consists of core-needle biopsies from early breast cancer patients. Athena slightly outperforms H-optimus-0 on the IMPRESS and ACROBAT datasets. Overall, Athena performs comparably to UNI but remains slightly below H-optimus-0 on average. The largest performance gap occurs in the HEROHE dataset, where H-optimus-0 clearly outperforms Athena.
Interestingly, some models underperform on specific datasets, such as Virchow on ACROBAT and Hibou-L on IMPRESS.

\begin{table}[h!]
\centering
\small
\setlength{\tabcolsep}{2pt}
\renewcommand{\arraystretch}{1.15}
\caption{HER2 classification AUC by model and dataset}
\label{tab:her2}
\begin{tabular}{l r r r r r r}
\toprule
\textbf{Model} & \scriptsize\textbf{YALE} & \scriptsize\textbf{HEROHE} & \scriptsize\textbf{IXORA} & \scriptsize\textbf{BCNB} & \scriptsize\textbf{IMPRESS} & \scriptsize\textbf{ACROBAT} \\
\midrule
Athena & 0.846 & 0.704 & 0.802 & 0.734 & \textbf{0.881} & \textbf{0.814} \\
Hibou-L & 0.732 & 0.719 & 0.705 & 0.715 & 0.545 & 0.775 \\
H-optimus-0 & \textbf{0.907} & \textbf{0.846} & \textbf{0.844} & 0.716 & 0.876 & 0.806 \\
Midnight-12 & 0.745 & 0.675 & 0.691 & 0.665 & 0.605 & 0.759 \\
UNI & 0.860 & 0.778 & 0.802 & \textbf{0.743} & 0.822 & 0.809 \\
Virchow & 0.851 & 0.753 & 0.659 & 0.659 & 0.765 & 0.599 \\
\bottomrule
\end{tabular}
\end{table}

\subsection{CAMELYON16}

For the CAMELYON16 dataset, the best-performing models are UNI, Athena, and H-Optimus, with H-Optimus achieving the highest overall performance. The weakest performance is observed for Hibou-L, as some training seeds produced poor results; since the reported score represents the average across all seeds, this reduced its final performance.

\begin{table}[h!]
\centering
\small                                % one step bigger
\setlength{\tabcolsep}{6pt}           % default-ish column spacing
\renewcommand{\arraystretch}{1.15}    % a bit taller rows
\caption{CAMELYON16 test AUCs for different FMs}
\label{tab:model-scores}
\begin{tabular}{l r}
\toprule
\textbf{Model} & \textbf{Score} \\
\midrule
H-optimus-0  & 0.9962 \\
UNI        & 0.9908 \\
Athena     & 0.9858 \\
Virchow    & 0.9642 \\
Midnight-12   & 0.9495 \\
Hibou-L      & 0.8449 \\
\bottomrule
\end{tabular}
\end{table}

\subsection{IDC vs. ILC Classification}

For the IDC/ILC classification task, the performances of the models are largely comparable. UNI and Virchow achieve the highest scores, followed by Athena and H-optimus, with Hibou-L and Midnight showing the lowest performance.

\begin{table}[h!]
\centering
\small                                % one step bigger
\setlength{\tabcolsep}{6pt}           % default-ish column spacing
\renewcommand{\arraystretch}{1.15}    % a bit taller rows
\caption{IDC vs ILC test AUCs for different FMs}
\label{tab:model-scores}
\begin{tabular}{l r}
\toprule
\textbf{Model} & \textbf{Score} \\
\midrule
UNI        & 0.9526 \\
Virchow    & 0.9521 \\
Athena     & 0.9485 \\
H-optimus-0  & 0.9471 \\
Hibou-L      & 0.9382 \\
Midnight-12   & 0.9270 \\

\bottomrule
\end{tabular}
\end{table}

\section{Discussion}

Across nearly all slide-level tasks, Athena outperforms Virchow and Hibou-L, achieving performance levels comparable to UNI and slightly below H-optimus-0. Notably, Athena was initialized from a model pretrained on natural images and trained with substantially fewer patch views. We attribute its strong performance to the high whole-slide diversity in our training dataset. This is consistent with evidence from a recent comprehensive survey that compared histopathology FMs across diverse downstream tasks and found that performance does not scale simply with the number of training images, but instead depends more critically on dataset diversity across sources, patient populations, and cancer types \cite{kather_survey}.

As seen in Table~\ref{tab:hest}, Athena ranks 2nd best in the HEST benchmark. However we observe large standard deviations across tasks, which often exceed the performance differences between models, limiting the benchmark’s discriminative power. Moreover, the patch-level tasks appear to be poor indicators of downstream performance. For example, UNI performs strongly on downstream tasks but ranks lowest on the HEST benchmark.
This suggests that new benchmarks are required to enable meaningful model comparisons. Until such benchmarks are available, model evaluations should focus on slide-level downstream tasks.

A key limitation of this work lies in the absence of a clear metric to quantify the diversity of our slide repository. It remains an open question which aspects of diversity are most relevant for model generalization: different color profiles, patient demographics, organ types, or scanner technologies. Future work should systematically investigate which forms of diversity are most critical for developing robust models with less data.

The initialization from a pretrained model worked well in our setup, indicating that its low-level features are sufficiently generalizable. However, training from scratch may enable the model to learn more domain-specific histopathology features and enhance performance, at the expense of longer training and greater data requirements.

Although our training was considerably less data- and compute-intensive than that of other recent models, it still required a setup of 32 H200 GPUs due to DINOv2’s large batch size requirements. This dependency on extensive GPU resources remains a key constraint in FM training.

Another limitation of our current approach is the use of completely random tissue sampling. This strategy may underrepresent rare but diagnostically relevant tissue structures while oversampling common regions with limited informational value. More advanced sampling strategies could help ensure a more balanced representation of tissue types and improve overall model robustness.

Previous studies have shown that pathology FMs often capture center-specific rather than diagnostic features \cite{unrobust}. In our case, the use of a diverse and geographically balanced training dataset should help mitigate this issue. Nonetheless, systematic evaluation across institutions and staining protocols remains essential to confirm that the FMs primarily rely on diagnostic, not confounding features.

\section{Conclusion}
%%\label{}
Our FM Athena approaches state-of-the-art performance and surpasses several models trained on substantially larger datasets. 
These results demonstrate that with sufficient diversity across institutions, scanners, and tissue types, powerful representations can be learned without billions of training patches, emphasizing that slide diversity, rather than raw patch quantity, drive model performance.

In conclusion, our findings demonstrate that slide diversity plays a more crucial role than the sheer number of training patches in developing histopathology FMs. By sampling moderate amounts of patches per slide from a diverse slide repository, models can achieve strong performance while requiring fewer computational and data resources.

Looking ahead, there remains considerable potential to further reduce the number of training patches required.
One promising direction is to adopt more advanced patch selection strategies, such as segmentation-based filtering, or feature-based clustering to ensure more representative sampling.
Additionally, stain-specific augmentations may help emulate dataset diversity, further lowering the data requirements.
This could free up resources to explore the impact of dataset composition, augmentations, and training hyperparameters, ultimately enabling the development of more robust FMs that generalize across cohorts, scanners, and staining conditions.

\section*{Data Availability}
The Athena model and pretrained weights are available at \href{https://huggingface.co/PAICON-GmbH/Athena-0}{https://huggingface.co/PAICON-GmbH/Athena-0}.

The training data used in this work consist primarily of proprietary histopathology slides that are not publicly available, except for the GTEx dataset~\cite{gtex}.
For the IDC/ILC task, publicly available slides from TCGA~\cite{tcga} were used.  
For CAMELYON16, the corresponding dataset is available~\cite{cam16}.  
For the MSI/MSS task, training slides were obtained from TCGA~\cite{tcga} and CPTAC~\cite{cptac}, while evaluation was performed using PAIP~\cite{paip}, SURGEN~\cite{surgen}, and NIB~\cite{nib}.  
For the HER2 task, all datasets except IXORA are publicly available: training data were taken from TCGA~\cite{tcga} and CPTAC~\cite{cptac}, and evaluation datasets included YALEHER2~\cite{yaleher2}, IMPRESS~\cite{impress}, BCNB~\cite{bcnb}, ACROBAT~\cite{acrobat}, and HEROHE~\cite{herohe}.

\section*{Author Contributions}

C.B. led concept development, methodology design, implementation, analyses, visualization, and wrote the original draft. 
M.R, J.W., M.Z. and B.A. contributed to data curation. 
M.P., W.A. and J.B. contributed to code development and implementation support. 
Sh.K. and R.K., Sw. K., B.K. and  S.V.C  assisted with dataset acquisition.
W.A. and C.A. supervised the work. 
All authors reviewed the manuscript.

\section*{Competing Interests}
C.B., J.W., M.Z., B.A., M.R., J.B., W.A. and C.A. are employees of PAICON GmbH. All other authors declare no competing interests.
%% \section*{Acknowledgements}
%% Thanks to ...


%% \section{Appendix title 1}

%% \section{Appendix title 2}

\bibliographystyle{elsarticle-num}
\bibliography{example}



%% For example:

%% \bibitem[Aladro et al.(2015)]{Aladro15} Aladro, R., Martín, S., Riquelme, D., et al. 2015, \aas, 579, A101

%%\end{thebibliography}

\end{document}